\documentclass[aps,prb,twocolumn]{revtex4}
\usepackage{bm,color,amsmath,amssymb,mathrsfs,latexsym,graphicx,psfrag}

\newcommand{\bk}{{\mathbf k}}

\newcommand{\br}{{\mathbf{r}}}
\newcommand{\be}{\begin{equation}}
\newcommand{\ee}{\end{equation}}

\def\bea{\begin{eqnarray}}
\def\eea{\end{eqnarray}}

\begin{document}
\title{Rotation Anomaly and Topological Crystalline Insulators}
\author{Chen Fang}
\email{cfang@iphy.ac.cn}
\affiliation{Beijing National Laboratory for Condensed Matter Physics and Institute of Physics, Chinese Academy of Sciences, Beijing 100190, China}
\author{Liang Fu}
\email{liangfu@mit.edu}
\affiliation{Department of Physics, Massachusetts Institute of Technology, Cambridge, MA 02139}

\date{\today}
\begin{abstract}
We show that in the presence of $n$-fold rotation symmetries and time-reversal symmetry, the number of fermion flavors must be a multiple of $2n$ ($n=2,3,4,6$) on two-dimensional lattices, a stronger version of the well-known fermion doubling theorem in the presence of only time-reversal symmetry. The violation of the multiplication theorems indicates anomalies, and may only occur on the surface of new classes of topological crystalline insulators. Put on a cylinder, these states have $n$ Dirac cones on the top and on the bottom surfaces, connected by $n$ helical edge modes on the side surface.
\end{abstract}
\maketitle

\section{Introduction}

\begin{figure*}
\begin{centering}
\includegraphics[width=1\linewidth]{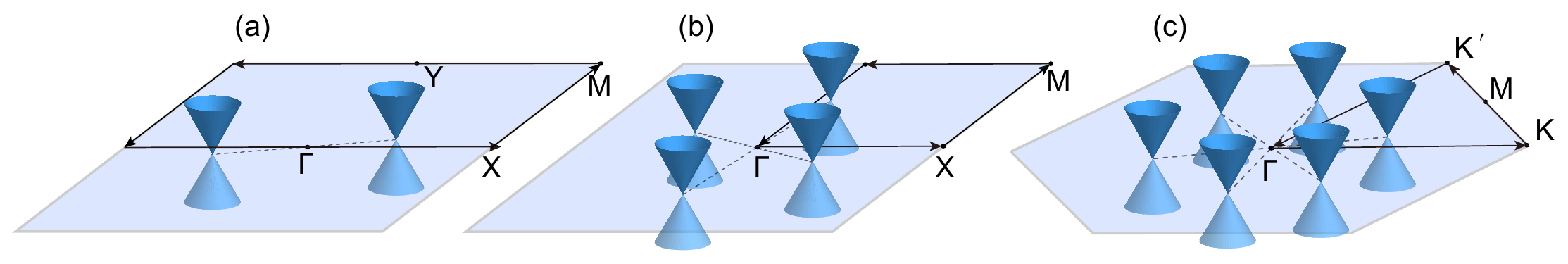}
\par\end{centering}
\protect\caption{\label{fig:1}The schematics of the gapless states in two dimensions that have rotation and time-reversal symmetries. There are (a) two, (b) four and (c) six Dirac cones, related to each other by two-, four- and six-fold rotation symmetries, respectively, in the first Brillouin zone. The contours are the boundaries of the invariant Brillouine zones, along which the Berry phase is quantized to either zero or $\pi$.}
\end{figure*}

A single flavor of massless relativistic fermion is known to have quantum anomalies, such that the conservation of certain global symmetry current is broken at the quantum level\cite{Bardeen1969}.
Well known examples include the chiral anomaly of Weyl fermions in three dimensions\cite{Adler1969,Bell1969,Nielson1981} and the parity anomaly in two dimensions\cite{Niemi1983,Haldane1988}.
Due to these anomalies, theories with only one flavor of massless fermions cannot have an ultraviolet(UV) completion that preserves the relevant symmetry.
This in turn implies that any UV-complete theory with $U(1)$ charge conservation must have an even number of massless fermions in odd spatial dimensions, and also in even spatial dimensions if time-reversal symmetry is present.

Massless relativistic fermions also emerge as low-energy excitations in solids, where the lattice spacing provides a natural ultraviolet cutoff.
The above fermion doubling theorem then enforces that Weyl fermions come in pairs in three-dimensional semimetals, and that Dirac fermions\footnote{Dirac fermions are massless two-component complex fermions in two dimensions, and are massless four-component complex fermions in three dimensions.} must come in pairs in two-dimensional (2D) time-reversal-invariant systems such as graphene.
Unlike the quantum vacuum in particle physics, condensed matter systems are non-Lorentz-invariant, but satisfy the spatial symmetry of the underlying crystal.
Spatial symmetry also constrains the number of massless fermions.
For example, recent works \cite{Fang2015,Shiozaki2015} have shown that any 2D lattice with either (glide) reflection symmetry, or the composite symmetry of twofold rotation and time-reversal, must have an even number of Dirac fermions.
The field theory of massless Dirac fermions in these systems is compatible with emergent Lorentz invariance at low energy.
Since the above symmetry transformations reverse the orientation of $2+1$-dimensional spacetime, their actions on emergent relativistic fermions are essentially equivalent to inversion symmetry in parity anomaly.

In this work, we present new quantum anomalies associated with time-reversal ($T$) and discrete rotational symmetry of crystals ($C_{n=2,4,6}$). These anomalies can only exist in theories without (emergent) Lorentz-invariance, and lead to a stronger constraint on the number of massless fermions in two dimensions. We show that in time-reversal-invariant band structures with $T^2=-1$ and $2m$-fold rotational symmetry, the number of stable massless Dirac fermions must be a multiple of $4m$.
Here massless Dirac fermion is synonymous with linear band crossing in Brillouin zone, and ``stable'' means that these band crossings are robust against arbitrary perturbations preserving $T$ and $C_{2m}$.
This result, dubbed the fermion multiplication theorem, is a generalization of fermion doubling theorem in particle physics to crystalline solids.

Despite lacking a simple lattice regularization, field theories with anomaly can describe boundary states of a topological bulk state in one higher spatial dimension.
This provides a powerful approach to identify and classify topologically nontrivial bulk states.
A single $3+1$ Weyl fermion with chiral anomaly appears on the boundary of a four-dimensional quantum Hall state \cite{Zhang2001}, and a single $(2+1)$-d massless Dirac fermion with parity anomaly appears on the surface of a three-dimensional topological insulator with time-reversal symmetry\cite{Fu2007}.
In both systems, the topological boundary states evade the fermion doubling theorem because their UV-completion lies in the bulk. This leads to a bulk-boundary correspondence: anomalous theory on the boundary imply nontrivial topology in the bulk. 

In the similar spirit, the new $2+1$D anomaly we found has important implications for three-dimensional topological insulators protected jointly by rotation and time-reversal symmetry, known as topological crystalline insulators\cite{Ando2015,Chiu2016,Shiozaki2017} (TCIs). The study of anomaly leads us to theoretically discover new classes of time-reversal-invariant TCIs with $C_{2m=2,4,6}$ rotation symmetry.
These TCIs have anomalous surface states on the top and bottom surfaces perpendicular to the $n$-fold axis. Such topological surface states consist of $2m$ Dirac cones that connect to bulk states at high energy, evading the above fermion multiplication theorem. Furthermore, these TCIs support $n$ one-dimensional helical modes on the side surface \cite{Song2017}.
For each class of new TCIs, we construct the corresponding $Z_2$ topological invariant in terms of Bloch wavefunctions in momentum space. We further provide a unified real-space understanding of these TCIs based on dimensional reduction and domain wall states. Finally, we predict several materials realizing the anomalous surface states protected by two- and four-fold rotation symmetries.

\section{Rotation anomaly}

We consider two-dimensional systems of non-interacting electrons with spin-orbit coupling, time-reversal and $n$-fold rotational symmetry, where $n=2,4$ or $6$. We aim to establish the number of symmetry-protected Dirac cones---or equivalently the number of stable band crossings---that such systems are allowed to have.
For this purpose, it suffices to consider systems with an integer number of electrons per unit cell and with either a band gap or Dirac points at Fermi energy.

Dirac points can be located at either generic momenta or high-symmetry points in the Brillouin zone. As shown in Appendix \ref{appn:1}, one can always choose a $C_n$ symmetric superlattice unit cell such that (1) high-symmetry points fold back to $\Gamma$ in the reduced Brillouin zone; (2) the number of electrons in the enlarged unit cell is an even integer. The condition of even-integer electron filling then guarantees that Dirac points at $\Gamma$ originating from Kramers degeneracy are away from Fermi energy. Hence, in the following analysis, we only study Dirac points at generic momenta, whose presence makes systems at certain even-integer electron fillings gapless.

Due to time reversal and $C_{n}$ rotation symmetry,  the number of Dirac points at generic momenta must be a multiple of $n$, each multiplet consisting of $n$ symmetry-related members at momenta $\pm \bk_1, \pm \bk_2, ... \pm \bk_{n/2} $, as shown in Fig.1.

Interestingly, we now show that the simplest scenario of $n$ Dirac points is anomalous,  i.e., it can only be realized on the surface of a  3D TCI. In any two-dimensional lattice system with time-reversal symmetry, the number of Dirac points away from high-symmetry points must be a multiple of $4$, $8$ or $12$ respectively, when 2-, 4-, or 6-fold rotation symmetry is present.

To prove this fermion multiplication theorem, consider the Berry phase along the closed contours in the Brillouin zone plotted in Fig.\ref{fig:1}.
Due to time-reversal and the two-fold rotation symmetry, the Berry curvature vanishes everywhere in momentum space, hence the Berry phase along each contour is quantized to be either $0$ or $\pi$\cite{Fang2015}.
In the latter case, there must be an odd number of Dirac cones enclosed by the contour.

Importantly, the Berry phases associate with the loops in Fig.\ref{fig:1}(a,b,c) are determined by the rotation eigenvalues at high-symmetry points $\Gamma, X, Y, M, K$ in the corresponding Brillouin zone. Following the method of Ref.\cite{Fang2012}, we find
\bea\nonumber
e^{i\Theta_2}&=&(-1)^{N_{occ}}\prod_{i=1,...,N_{occ}}\zeta_i(\Gamma)\zeta_i(X)\zeta_i(Y)\zeta_i(M),\\
e^{i\Theta_4}&=&(-1)^{N_{occ}}\prod_{i=1,...,N_{occ}}\xi_i(\Gamma)\xi_i(M)\zeta_i(X),\\\nonumber
e^{i\Theta_6}&=&(-1)^{N_{occ}}\prod_{i=1,...,N_{occ}}\omega_i(\Gamma)\theta_i(K)\zeta_i(M), 
\label{theorem}
\eea
where $N_{occ}$ is the number of occupied bands (counting spin),
 $\zeta_i, \theta_i, \xi_i$ and $\omega_i$ are the eigenvalues of $C_{2},C_3,C_4,C_6$ operation at the corresponding fixed points.

Since we deal with even-integer electron fillings, $N_{occ}$ is even.
Due to time-reversal, at $\Gamma, X, Y, M$ points, rotation eigenvalues $\zeta, \xi, \omega$ appear in complex conjugate pairs, leading to Kramers degenerate bands. It then follows from (\ref{theorem}) that $e^{\Theta_{2}} = e^{i \Theta_4}=1$.
For systems with $C_6$ symmetry, at $K$ point, the allowed $C_3$-eigenvalues are $e^{i2\pi/3}$, $e^{-i2\pi/3}$ or $-1$. The first two appear in complex conjugate pairs due to the composite symmetry $C_2T$.  The number of the remaining bands with $C_3$-eigenvalue $\theta=-1$ must also be even, as the total number of occupied bands is even. This implies $e^{\Theta_{6}}=1$. To summarize, we find 
\bea\label{eq:theorem}
\Theta_{2,4,6}=0\;\textrm{mod}\;2\pi. 
\eea

Eq.(\ref{eq:theorem}) implies that within each contour, there must be even number of Dirac cones.
In other words, all configurations in Fig.\ref{fig:1} are anomalous, because they all possess one Dirac cone enclosed by the contour.
These states can only be realized on the surface of  topological crystalline insulators, to which we turn our attention below.

\section{New topological crystalline insulators: momentum space}

Topological crystalline insulators previously considered have Dirac cones pinned to high-symmetry points\cite{Fu2011,Liu2014,Alexandradinata2014,Wieder2017} or lines\cite{Fang2015,Shiozaki2015,Wang2016,Chang2017} (but see a singular exception in Ref.[\onlinecite{Shiozaki2014,Fang2015}]); therefore the surface states in Fig.\ref{fig:1} where the Dirac cones are at generic momenta must belong to new types undiscovered.
We show that they can all be understood as the superpositions of two topological insulators.
Here we start by considering the case of twofold rotation.
With twofold rotation and time-reversal, the surface state Hamiltonian reads (up to some rescaling and rotation of frame of reference)
\begin{equation}
h_\pm(\mathbf{k})=k_x\sigma_x\pm{k}_y\sigma_y,
\end{equation}
where $\sigma_{i=x,y,z}$ are the Pauli matrices, and $\pm$ is the helicity of the surface states.
The twofold rotation and the time-reversal symmetry are represented by $C_2=i\sigma_z$ and $T=(i\sigma_y)K$.
Consider the superposition of $h_+(\mathbf{k})$ and $h_-(\mathbf{k})$
\begin{equation}
H(\mathbf{k})=h_+(\mathbf{k})\oplus{h}_-(\mathbf{k})=k_x\tau_0\sigma_x+k_y\tau_z\sigma_y,
\end{equation}
where $\tau_i$ are Pauli matrices acting in the flavor space.
One can check that no `mass term' can be written to gap the dispersion of $H(\mathbf{k})$ if $C_2$ and $T$ were to be preserved.
In fact, the only $T$ invariant mass term is proportional to $\tau_y\sigma_y$, which breaks $C_2$. Symmetric terms only split the two Dirac cones at $\mathbf{k}=0$ to two, as shown in Fig.\ref{fig:1}(a).
This shows that the new topological crystalline insulators' surface state in Fig.\ref{fig:1}(a) can be realized by superimposing two topological insulators whose surface states have opposite helicity.
Since one can show that two surface states with the same helicity can be symmetrically gapped out with a mass term proportional to $\tau_y\sigma_z$, we conclude that with twofold rotation and time-reversal, the classification is $\mathbb{Z}_2\times\mathbb{Z}_2$, where the generator of each $\mathbb{Z}_2$ is a topological insulator whose surface states have $+1$ or $-1$ helicity.
In terms of this topological index, the surface states shown in Fig.\ref{fig:1}(a) belong to topological crystalline insulator with index $(11)$. In Table.\ref{tab:1}, we show how to construct the surface states in Fig.\ref{fig:1}(b,c) from superimposing two topological insulators that have fourfold and sixfold rotation symmetries. The proof of these results is found in Appendix \ref{appn:2}.
\begin{table}
\begin{centering}
\includegraphics[width=1\linewidth]{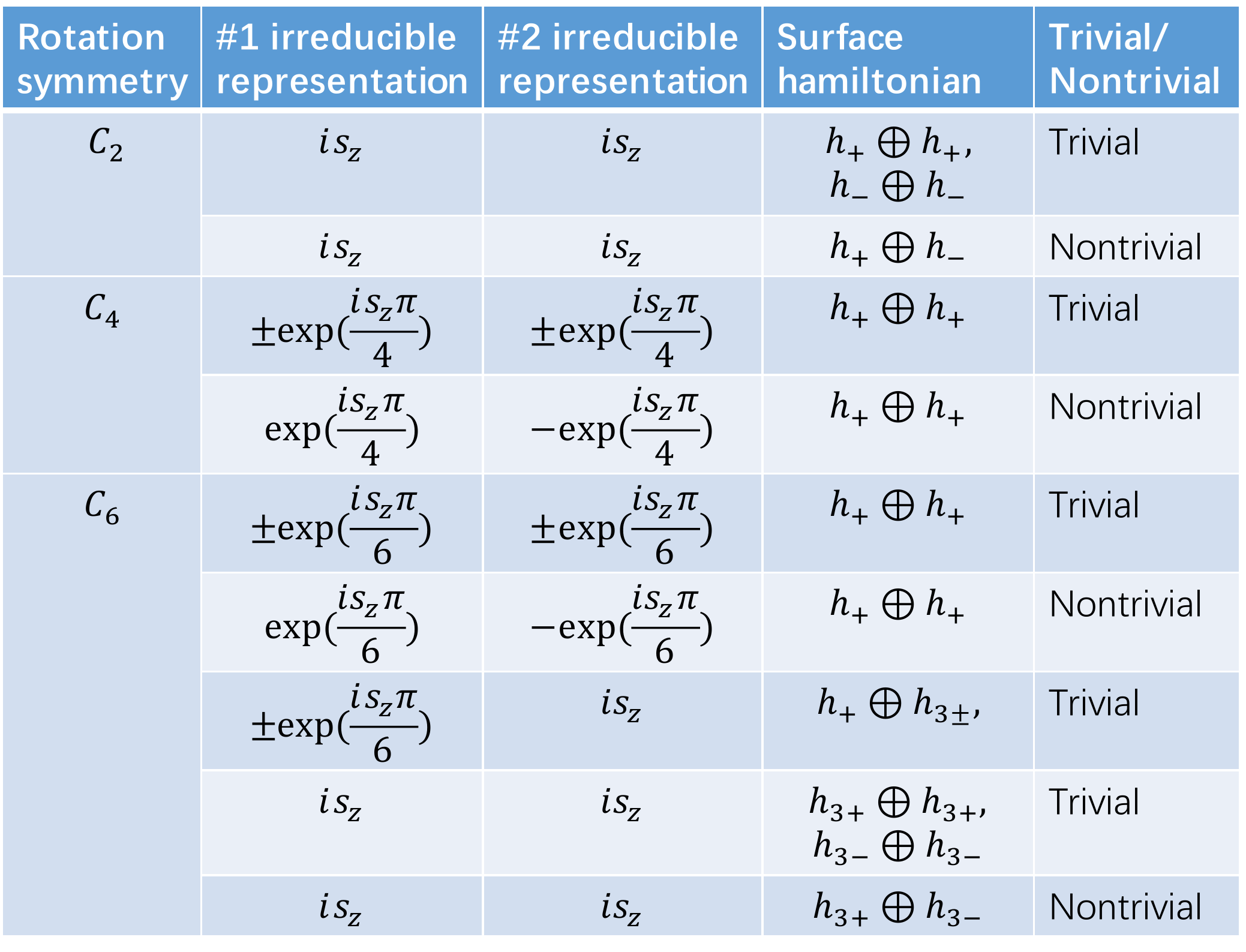}
\par\end{centering}
\protect\caption{\label{tab:1}Superposition of the surface hamiltonians of two topological insulators that have two-, four- and sixfold rotation symmetries, and time-reversal symmetry represented by $is_yK$. This table lists all possible combinations of irreducible representations at the Dirac points as well as the explicit forms of each surface hamiltonian compatible with the representations, where $h_\pm(k_x,k_y)\equiv{k}_xs_x+k_ys_y$ and $h_{3\pm}(k_x,k_y)\equiv(k_x+ik_y)^3(s_x\pm{i}s_y)+h.c.$. In the last column, ``trivial'' and ``nontrivial'' means that the surface states can and cannot be symmetrically gapped, respectively.}
\end{table}

\section{New topological crystalline insulators: real space}

\begin{figure}
\begin{centering}
\includegraphics[width=1\linewidth]{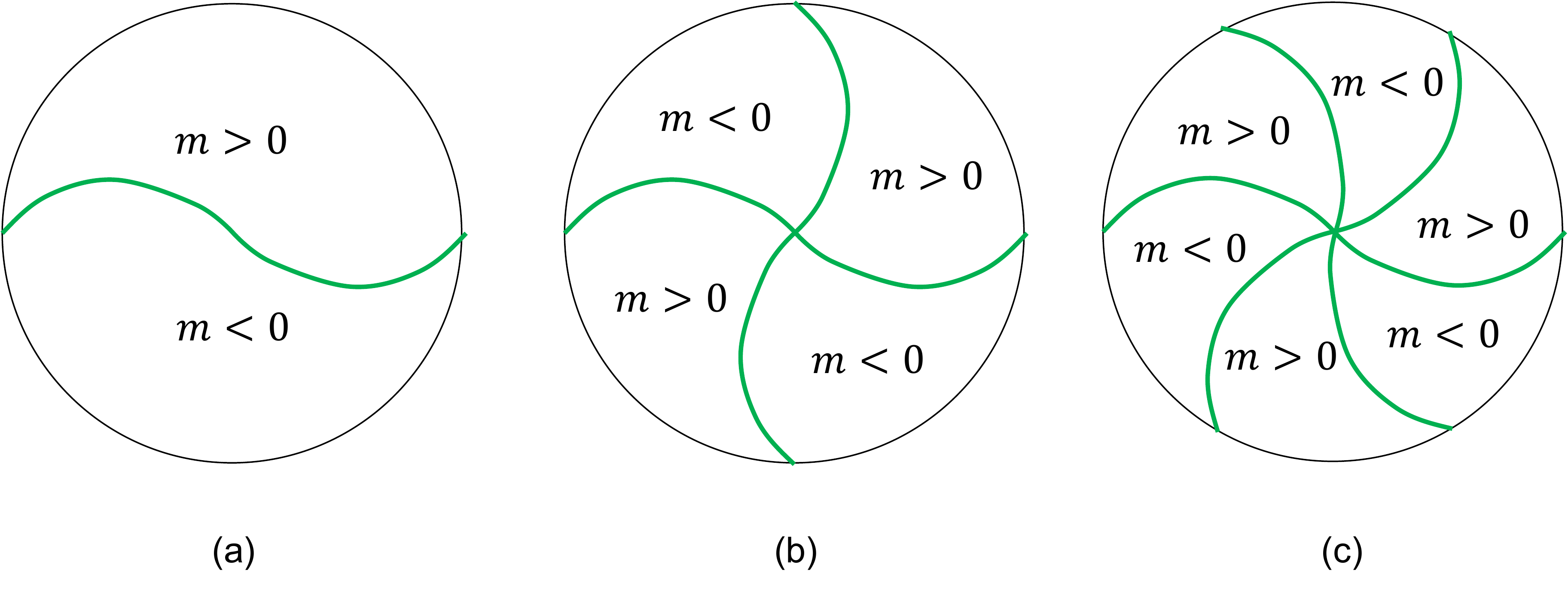}
\par\end{centering}
\protect\caption{\label{fig:2}Real space mass distribution on the top (bottom) surface having (a) twofold, (b) threefold and (c) sixfold rotations and time-reversal. The mass field gaps out most regions of the surface, but leaving one, two and three domain walls (green curves) due to mass inversion enforced by rotation symmetry.}
\end{figure}

One can also understand the nontrivial nature of the surface state from the perspective of real space, again starting from the case of twofold rotation.
Consider a generic point away from the origin on the surface, where the only symmetry in its neighborhood is time-reversal.
One can add the time-reversal-invariant mass term $m\sigma_y\sigma_y$, and the surface states in that region are gapped.
But since $\tau_y\sigma_y$ and $C_2=i\sigma_z$ anticommute, the mass term in real space changes sign after a twofold rotation, or $m(\br)=-m(-\br)$ for any $\br$.
Any real function $m(\br)$ satisfying this equation always has a nodal line where $m(\br)=0$, passing through the origin. The nodal line can be considered a domain wall of mass, along which runs a helical edge mode [see Fig.\ref{fig:2}(a)].
As long as twofold rotation and time-reversal are preserved, the domain wall exists, while its shape can be arbitrarily deformed.
For fourfold and sixfold rotations, one can prove that the mass term satisfies $m(\br)=-m(C_{4,6}\br)$, which enforces two and three domain walls with helical modes [see Fig.\ref{fig:2}(b,c)].
These helical modes are the boundary manifestation of the nontrivial topology of the bulk on the top (bottom) surface in a slab geometry in the absence of translation symmetry.

\begin{figure}
\begin{centering}
\includegraphics[width=1\linewidth]{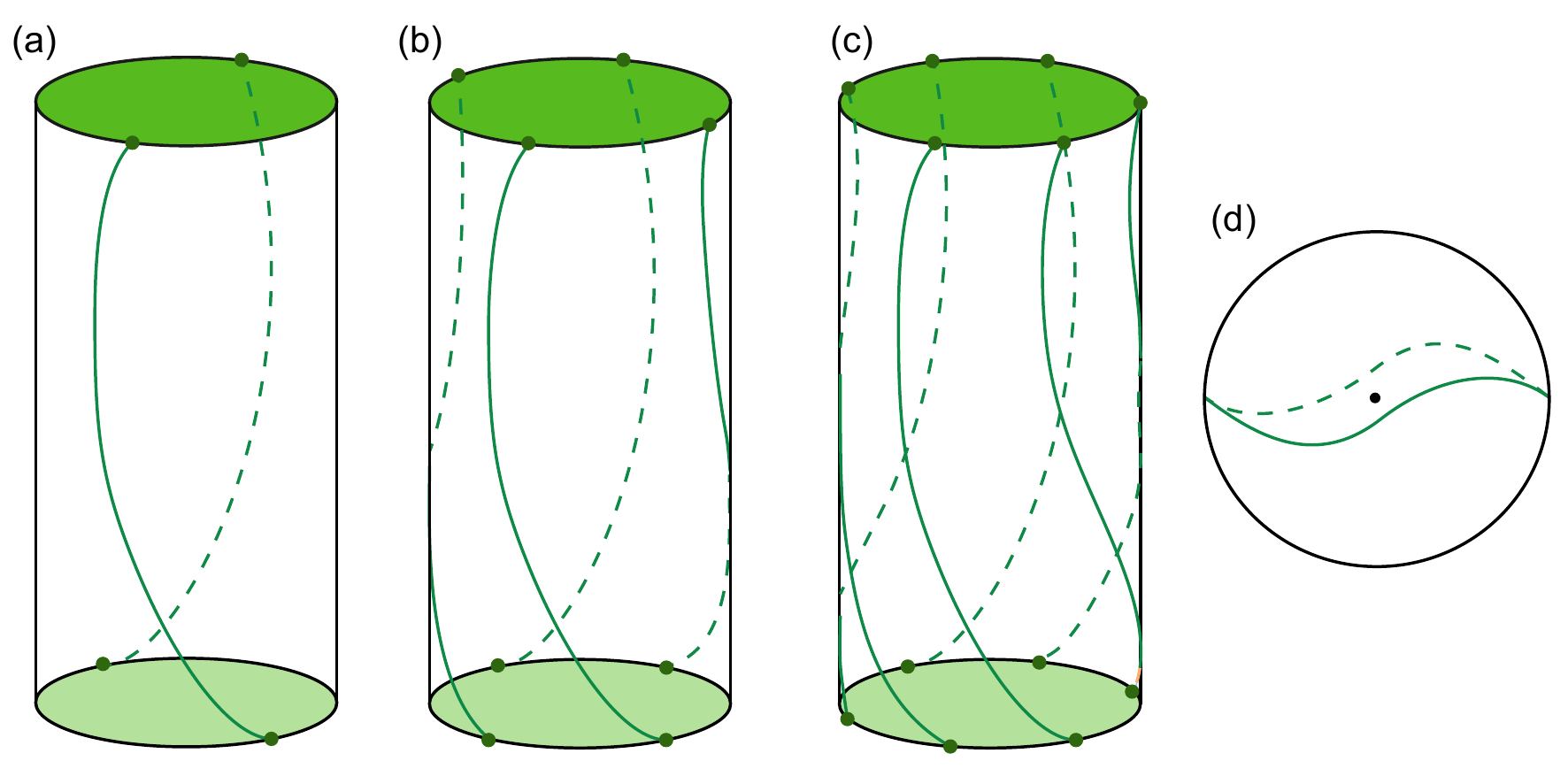}
\par\end{centering}
\protect\caption{\label{fig:3}Schematics of surface states on the top and the bottom surfaces and the edge states on the otherwise gapped side surfaces of the new topological crystalline insulators protected by (a) twofold, (b) fourfold and (c) sixfold rotation symmetries in cylinder geometry. The top and the bottom surfaces have Dirac cones shown in Fig.\ref{fig:1}(a,b,c), and on the side surface, two, four and six helical edge modes connect the two surfaces; they can have arbitrary shape and position but are related to each other by twofold, fourfold and sixfold rotations, respectively. In (d), an inversion symmetry protected topological crystalline insulator was put on a sphere, having a helical mode on a curve made from pairs of antipodal points, cutting the surface into two exact halves.}
\end{figure}

To gain further understanding of the bulk-edge correspondence in the new topological crystalline insulators, let us consider the states put on a cylinder with open side surfaces.
Since both the top and the bottom surfaces are anomalous, there must be gapless modes connecting them on the side surface.
These modes are the one-dimensional helical edge modes discussed in Ref.[\onlinecite{Song2017}], where the case of fourfold rotation was presented in detail.
For topological crystalline insulators protected by $C_{2,4,6}$, on the side surface there are exactly two, four and six helical edge modes connecting the states on the top and the bottom surfaces, schematically shown in Fig.\ref{fig:3}(a,b,c).
The existence of the edge modes can be intuitively understood starting from the domain wall picture presented above: the mass domains can be extended from the top (bottom) surface to the side surface, so that the helical modes along the domain walls also flow to the side surface; the number of helical edge modes equals twice the number of the domain walls.
Up to this point, we have shown that the boundary manifestation of the new topological crystalline insulators is twofold: on the top and the bottom surfaces, there are surface states that have $n$ Dirac cones; and on the side, there are $n$ one-dimensional helical modes connecting the two surfaces.
We remark that the only constraint placed by symmetry on these edge modes is that they are related to each other by $C_{n=2,4,6}$, but not pinned to any physical hinges or intersections of crystalline surfaces like in Ref.[\onlinecite{Benalcazar2017,Schindler2017,Langbehn2017,Benalcazar2017a}].
The existence of the one-dimensional edge modes hints us to write down the bulk topological invariants following Ref.[\onlinecite{Song2017}], and the details are given in Appendix \ref{appn:3}.
In Appendix \ref{appn:4}, we in addition discuss, as contrast, a related topological crystalline insulator protected by inversion and time-reversal\cite{Turner2010,Hughes2011}, where surface states do not exist, but a single helical edge mode exists on the surface [see Fig.\ref{fig:3}(d)].

Most three-dimensional topological crystalline states can be understood from a dimensional reduction perspective\cite{Isobe2015,Ezawa2016,Huang2017}, where the 3D state can be considered a set of decoupled layers of 2D topological states. From this perspective, all three types of new topological crystalline insulators in this paper can also be constructed from 2D topological insulators. This construction helps enhance the theory to include strongly interacting symmetry protected topological states protected by rotation symmetry and any local symmetry, including but not limited to time-reversal. The discussion is found in Appendix \ref{appn:5}.

\section{Realization of anomalous surface states in materials}

Finally, we propose materials where the anomalous surface states in Fig.\ref{fig:1} (a,b) can be observed, where there are two and four Dirac cones on a surface having twofold and fourfold rotation, respectively.
In fact, these two types of surface states have been predicted in SnTe, on the $(110)$-surface and the $(001)$-surface, respectively.
On the $(110)$-surface, there are two Dirac cones located along $\bar\Gamma\bar{Z}$ in the surface Brillouin zone, which have been considered protected by mirror symmetry\cite{Liu2013}.
But now we know that even when mirror symmetry is broken (via external strain field for example), as long as $C_2$ and $T$ are preserved, this surface state is still \emph{anomalous} and cannot be gapped.
Similarly, the four Dirac cones on the $(001)$-surface have also been considered protected by mirror symmetry\cite{Hsieh2012}, and we now emphasize that they are stable against all $C_4$ and $T$ preserving perturbations because of the $C_4$-rotation anomaly.
When mirror symmetries are broken, the Dirac points are free to move away from the high-symmetry lines to generic momenta on the $(110)$- and $(001)$-surfaces, as shown in Fig.\ref{fig:1}(a,b), respectively.

Another line of thinking starts from band inversion in the bulk, and from the observation that the new state can be constructed from superimposing two topological insulators.
In the case of $C_4$, it requires that the two surface Dirac points from the two topological insulators belong to different irreducible representation of $C_4$.
In the bulk, this means that two valence bands and two conduction bands have band inversions, and that the two valence bands (as well as conduction bands) belong to different irreducible representations of $C_4$.
One possible way of realization is through the inversion of two $\Gamma_8$ bands that have opposite parity.
$\Gamma_8$ is a four-dimensional irreducible representation, consisting of two Kramers' doublets with total angular momentum $j_z=\pm1/2\hbar$ and $j_z=\pm3/2\hbar$, respectively.
Under $C_4$, the two doublets transform differently as $\exp(i\sigma_z\pi/4)$ and $-\exp(i\sigma_z\pi/4)$, that is, differently.
Band inversion between two $\Gamma_8$ bands having opposite parity have been proposed in anti-perovskite material Sr$_3$PbO\cite{Hsieh2014}.
On the (001)-surface of Sr$_3$PbO, according to our theory, there will be four Dirac cones as shown in Fig.\ref{fig:1}(b).

On the side surface, the localized one-dimensional helical edge modes require the absence of gapless surface modes, but in all the above examples, the additional mirror symmetries of the lattice makes low-index surfaces such as $(100)$ and $(210)$ gapless.
In order to see the helical edge modes on the side, one has to (i) either break mirror symmetries by external perturbation such as strain or (ii) look at a high-index plane (e.g., $(221)$-surface) and its $C_4$ equivalents.

\acknowledgements{C.F. was supported by the National Key Research and Development Program of China under grant No. 2016YFA0302400, by NSFC under grant No. 11674370. L.F. is supported by DOE Office of Basic Energy Sciences, Division of Materials Sciences and Engineering under Award DE-SC0010526.}

\onecolumngrid

\begin{appendix}
\section{All Dirac cones can be moved to generic momenta with symmetric perturbations}
\label{appn:1}
We consider the Dirac cones appearing between the $\nu$-th and the $\nu+1$-th bands in some 2D system. We assume that the total number of Dirac cones is even, because if it is odd we return to the well-known surface state of a topological insulator protected by time-reversal symmetry. When the total number is even, there must be an even of Dirac cones at generic momenta and an even number of Dirac cones at high-symmetry momenta. Here these points are: $\Gamma$, $X$, $Y$, $M$ in a $C_2$-symmetric lattice, $\Gamma$, $X$ and $M$ in a $C_4$-symmetric lattice, and $\Gamma$, $K$ and $M$ in a $C_6$-symmetric lattice. Now we show that one can break translation symmetry without breaking rotation or time-reversal, such that all the original high-symmetry points are folded to the origin, i. e., $\Gamma$ in the folded Brillouin zone.
For the $C_2$-case, simply add the following charge density wave perturbation
\bea
\delta{V}(\br)=V_0[\cos(\mathbf{X}\cdot\br)+\cos(\mathbf{Y}\cdot\br)],
\eea
and one can easily check that $\mathbf{X}$, $\mathbf{Y}$ and $\mathbf{M}$ are folded back to $\Gamma$.

For the $C_4$-case, there are two $X$-points, denoted by $X_1$ and $X_2$, and we consider the perturbation
\bea
\delta{V}(\br)=V_0[\cos(\mathbf{X_1}\cdot\br)+\cos(\mathbf{X_2}\cdot\br)],
\eea
so that $\mathbf{X_1}$, $\mathbf{X_2}$ and $\mathbf{M}$ are all folded to $\Gamma$.

Finally, for the $C_6$-case, we consider the following perturbation
\bea
\delta{V}(\br)=V_1[\cos(\mathbf{K}\cdot\br)+\cos(\mathbf{K}\cdot{C}_3\br)+\cos(\mathbf{K}\cdot{C}_3^{-1}\br)]+V_2[\cos(\mathbf{M}_1\cdot\br)+\cos(\mathbf{M}_2\cdot\br)+\cos(\mathbf{M}_3\cdot\br)],
\eea
and $\mathbf{K}$, $-\mathbf{K}$, $\mathbf{M}_{1,2,3}$ are all folded to $\Gamma$.

It is easy to explicitly check that these terms preserve rotation and time-reversal. After these translation-breaking perturbations, there are an even number of Dirac cones at $\Gamma$. Each Dirac point is an irreducible representation of the symmetry group. Shifting their relative energy perturbatively preserves the rotation and time-reversal symmetries. Observe that starting with any filling number $\nu$, after folding the new filling number $\nu=4\nu$ for $C_2$- and $C_4$-cases and $\nu'=6\nu$ for the $C_6$-case, so that $\nu'$ is always even. For any even integer filling, it is always possible to remove the Dirac points away from $\Gamma$ to generic momenta by infinitesimally shifting their energies.

Up to this point, we have shown that all Dirac cones can be perturbatively moved to generic momenta without symmetry-breaking for any given Fermi energy (filling), as long as the total number of Dirac cones is even.

\section{Anomalous surface states protected by $C_{4,6}$-symmetry and time-reversal}
\label{appn:2}

We show how the anomalous surface states in Fig.1(b,c) can be constructed from two Dirac cones from $\Gamma$, belonging to two different representations of $C_4$ and $C_6$, respectively.
First look at the case of $C_4$. In the presence of time-reversal and the absence of SU(2) spin rotation symmetry, the symmetry group is then generated by $C_4$ and $T$ satisfying $C_4^4=-1, T^2=-1$ and $[C_4,T]=0$. These conditions give two distinct irreducible represetations
\bea
C_{4\pm}&=&\pm{e}^{is_z\pi/4},\\
T&=&is_yK.
\eea
Consider putting these two representations together one has a reducible four-dimensional representation
\bea
C_4&=&\tau_ze^{is_z\pi/4},\\
T&=&i\tau_0s_yK.
\eea
The minimal 2D theory compatible with this symmetry group is
\bea
h(k_x,k_y)=k\cos\theta\tau_0s_x+k\sin\theta\tau_0s_y,
\eea
where $(k,\theta)$ are the polar coordinates of $\mathbf{k}$.
The only time-reversal invariant mass term is proportional to $\tau_ys_z$, but this matrix anticommutes with $C_4$. Therefore the theory cannot be symmetrically gapped. We can add symmetry allowed terms
\bea
\delta{h}(k_x,k_y)=m_1\tau_zs_0+m_2k^2\tau_x\sin[2(\theta-\alpha)]s_0,
\eea
where $\alpha$ is some arbitrary number.
The dispersion is gapless at $k=|m_1|$ and $\theta=\alpha,\alpha+\pi/2,\alpha+\pi,\alpha+3\pi/2$, that is, four points as shown in Fig.1(b).

Then we look at the case of $C_6$ and $T$, satisfying $C_6^6=T^2=-1$ and $[C_6,T]=0$. There are three distinct irreducible representations of this symmetry group
\bea
C_6&=&\pm{}e^{is_z\pi/6}, is_z\\
T&=&is_yK.
\eea
For $C_6={e}^{is_z\pi/6}$, the minimal Dirac theory is
\bea
h_1(k,\theta)=k\cos\theta{s}_x+k\sin\theta{s}_y,\\
\eea
and for $C_6=-e^{is_z\pi/6}$, the minimal theory is the same
\bea
h_2(k,\theta)=h_1(k,\theta).
\eea
For $C_6=is_z$, the minimal theory is
\bea
h_{3\pm}=k^3\cos3\theta{s}_x\pm{k}^3\sin3\theta{s}_y.
\eea
Now we consider superimposing two of the four different theories, and ask if the resultant theory can be symmetrically gapped. We first consider the superposition of $h_1$ and $h_3+$,
\bea
H_{1,3+}\equiv{}h_1\oplus{h}_3.
\eea
We can add the following terms
\bea
\delta{H_{1,3+}}=m_1\tau_zs_0+km_2(\cos\theta\tau_xs_z+\sin\theta\tau_y),
\eea
and the dispersion becomes fully gapped and also $\theta$-independent
\bea
E(k)=\pm\sqrt{(k\pm{m}_1)^2+m_2^2k^2}.
\eea
Similarly, we can find that $H_{1,3+}\equiv{h}_1\oplus{h}_{3-}$ and $H_{2,3\pm}\equiv{h}_2\oplus{h}_{3\pm}$ can also be symmetrically gapped.
Now we focus on $H_{1,2}\equiv{h}_1\oplus{h}_2$, and add the following terms
\bea
\delta{H}_{1,2}=m_1\tau_z{s}_0+m_2k^3\sin3(\theta-\alpha)\tau_xs_z
\eea
to obtain the dispersion
\bea
E_{1,2}(k,\theta)=\pm\sqrt{k^2+m_1^2+m_2^2k^6\sin^2[3(\theta-\alpha)]}.
\eea
The dispersion has six Dirac points at $k=|m_1|$, $\theta=m\pi/3+\alpha$ for $m=0,1,2,3,4,5$, arranged in a configuration shown in Fig.1(c).
Finally, we study the case
\bea
H_{3+,3-}(k,\theta)\equiv{}k^3\cos3\theta\tau_0s_x+k^3\sin3\theta\tau_0s_y,
\eea
and we add terms
\bea
\delta{H}_{3+,3-}(k,\theta)=m_1\tau_zs_0+m_2\sin3(\theta-\alpha)\tau_xs_y.
\eea
There are six Dirac points at the same positions as in the case of $H_{1,2}$.
To summarize, for the $C_6$-case, the anomalous configuration of Dirac cones shown in Fig.1(c) can be realized by superimposing two combinations of two Dirac theories: (i) $h_1$ and $h_2$ and (ii) $h_{3+}$ and $h_{3-}$, while all other combinations can be gapped by symmetric perturbations.

\section{Bulk invariants}
\label{appn:3}

In this section, we propose the forms of the bulk invariants which characterize the new TCI having topological surface states shown in Fig.1(a,b,c), following the method outlined in Ref.[\onlinecite{Song2017}]. These invariants are expressed in terms of the $Z_2$-flow of Wannier centers (WCs) between the $k_z=0$ and the $k_z=\pi$ planes, considering each $k_z$-slice as a 2D subsystem. 

The Wannier centers are defined in the following way. For a given gapped 2D Hamiltonian $H(k_x,k_y)$ with occupied bands whose periodic parts of the wavefunctions are denoted by $|u_i(k_x,k_y)\rangle$, where $i=1,...,\nu$. In general, $|u_i(k_x,k_y)\rangle$ is not a continuous function of $\bk$. If the Chern number for all occupied bands vanish, it is proved that one can always find another set of functions $|w_i(k_x,k_y)\rangle$ that are (i) smooth for $\bk$ on a two-torus and (ii) the following equation
\bea
\sum_{i=1,...,\nu}|w_i(\bk)\rangle\langle{w}_i(\bk)|=\sum_{i=1,...,\nu}|u_i(\bk)\rangle\langle{u}_i(\bk)|
\eea
is satisfied at each $\bk$. Wannier centers are two-vectors $(P_{ix},P_{iy})$, defined as the 2D-polarization of all $|w_i(\bk)\rangle$\cite{Fang2012}
\bea\nonumber
P_{ix}&\equiv&\int\frac{dk^2}{4\pi^2}i\langle{w}_i(\bk)|\partial_{k_x}|w_i(\bk)\rangle,\\
P_{iy}&\equiv&\int\frac{dk^2}{4\pi^2}i\langle{w}_i(\bk)|\partial_{k_y}|w_i(\bk)\rangle.
\eea
Physically, the WCs can be considered the ``charge centers'' of the electronic bands, and give us real space understanding of the electronic structure of an insulator. It is obvious that $|w'_i(\bk)\rangle=U_{ij}(\bk)|w_j(\bk)\rangle$ is another set of functions satisfying conditions (i) and (ii) above for any unitary $U(\bk)$; therefore, WCs are highly gauge-invariant quantities. Unlike in 1D, there is no ``canonical'' way of choosing $|w_i(\bk)\rangle$'s whose Fourier transform give maximally localized Wannier functions\cite{Marzari1997}.

Thanks to the insight in Ref.[Soluyanov2011,Po2017,Bradlyn2017], we now understand how WCs are related to nontrivial topology of the system: one can (cannot) find a set of $|w_i(\bk)\rangle$ that forms a representation of the space group when the insulator is topologically trivial (nontrivial). For example, when the insulator having two occupied bands is a 2D TI, the $|w_i(\bk)\rangle$'s will \emph{never} satisfy $T|w_i(\bk)\rangle=\epsilon_{ij}|w_j(-\bk)\rangle$, or that the two WCs do \emph{not} coincide. In the following, we make the following conjecture: in 2D with time-reversal, rotation symmetry does \emph{not} give us additional topologically nontrivial classes. It is equivalent to the statement that when a 2D time-reversal invariant insulator is not a time-reversal TI, it is equivalent to some atomic insulator. For an atomic insulator, one can always find a set of $|w_i(\bk)\rangle$ that form a double-valued representation of the space group. In the following, we have assumed that $k_z=0$-slice and $k_z=\pi$-slice are not 2D time-reversal TIs, so that each $k_z$-slice is topologically equivalent to a 2D atomic insulator. This conjecture is supported by the absence of symmetry indicators when only time-reversal and rotation symmetries are present in 2D in class AII\cite{Po2017}.

\begin{figure}
\begin{centering}
\includegraphics[width=1\linewidth]{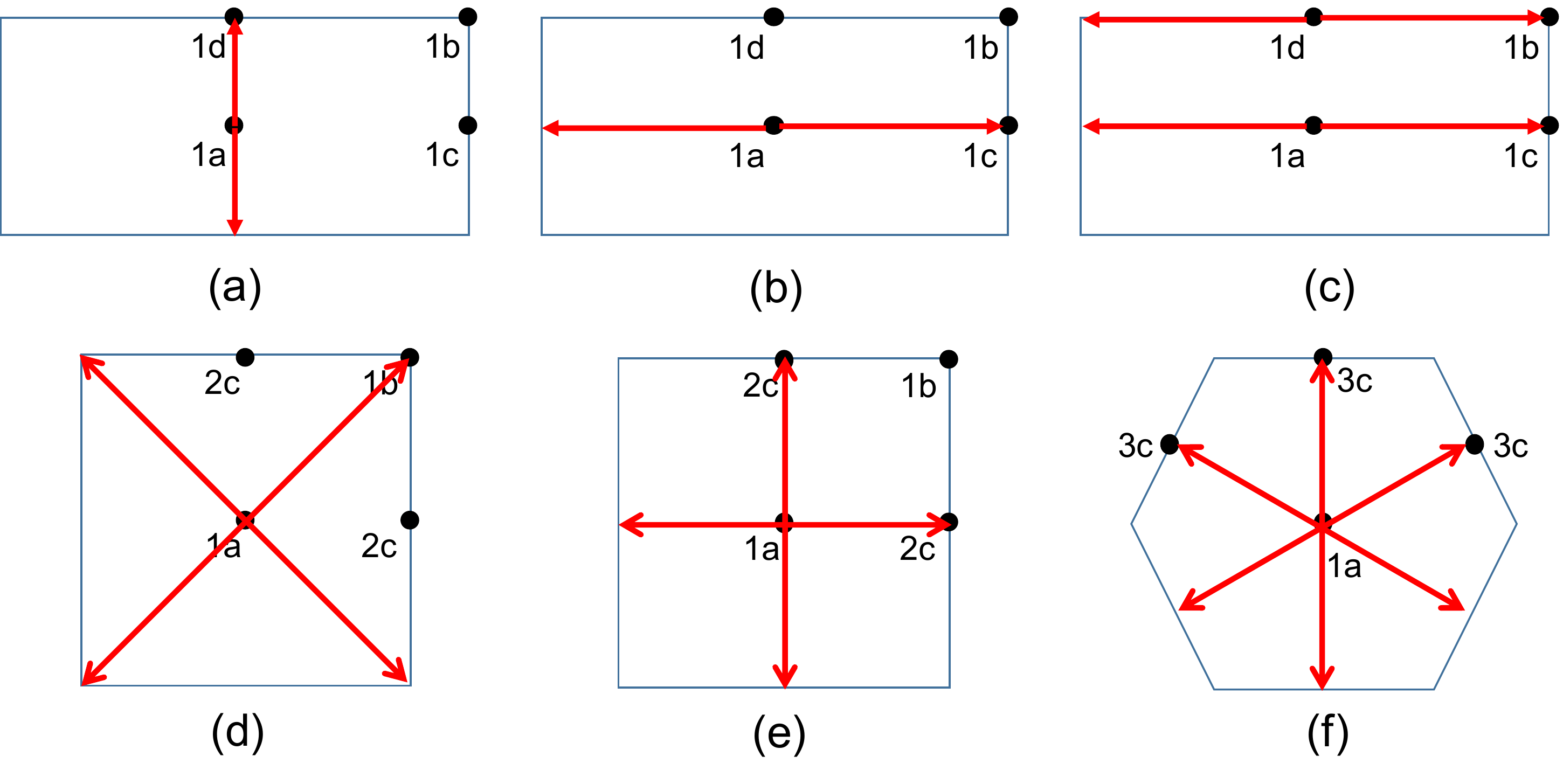}
\par\end{centering}
\protect\caption{\label{fig:S1}The definition of Wyckoff positions in simple rectangular, square and triangular lattices and the nontrivial flows of Wannier centers as $k_z$ changes from $0$ to $\pi$.}
\end{figure}

While WCs are in general gauge-variant, in Ref.[\onlinecite{Song2017}], it has been shown that at $k_z=0$ and $k_z=\pi$, in the presence of $C_4$ and time-reversal, the numbers of WCs at Wyckoff positions $1a$ and $1b$ modulo eight are gauge-invariant. Similarly, one can prove that following: (i) in the presence of $C_2$ and $T$, the numbers of WCs at 1a, 1b, 1c and 1d modulo four are gauge-invariant; (ii) in the presence of $C_4$ and $T$, the numbers of WCs at 1a and 1b modulo eight are gauge invariant, and the number of WCs at 2c modulo four is gauge invariant; (iii) in the presence of $C_6$ and $T$, the number of WCs at 1a modulo twelve is gauge-invariant and the number of WCs at 3c modulo four is gauge invariant. For the definitions of the Wyckoff positions, see Fig.\ref{fig:S1}.
Away from these two high-symmetry slices at $k_z=0$ and $k_z=\pi$, the WCs do not have to appear in Kramers' pairs, but their distribution should be symmetric under $C_{2,4,6}$. Their positions are, however, completely gauge-variant. If one plots the WCs as a function of $k_z$, one obtains a set of curves connecting the WCs at $k_z=0$ and $k_z=\pi$, that is, flow between these two sets of WCs. Physically, these flows indicate how the charge center moves as $k_z$ goes from $0$ to $\pi$, and therefore tell us if surface/edge states appear when we vertically cut the insulator\cite{Song2017}. A typical trivial flow is zero flow when the WCs at $k_z=0$ and at $k_z=\pi$ are the same, and we define trivial flow as all flows that can be adiabatically tuned to a trivial flow.

The nontrivial flows in the $C_{2,4,6}$ cases are plotted in Fig.\ref{fig:S1}. The flows in Fig.\ref{fig:S1}(a,b,e) corresponds to weak topological insulators, and the flows in Fig.\ref{fig:S1}(c,d,f) correspond to the new topological insulators studied in the main text, with the new $Z_2$-index. For the $C_{2,4,6}$-cases, the new $Z_2$-indices are nontrivial if and only if
\bea
N^{0}_{WC}(\Gamma,X,Y,M)-N^\pi_{WC}(\Gamma,X,Y,M)&=&2\;\mathrm{mod}\;4,\\
N^{0}_{WC}(\Gamma,M)-N^\pi_{WC}(\Gamma,M)&=&4\;\mathrm{mod}\;8,\\
N^{0}_{WC}(\Gamma)-N^\pi_{WC}(\Gamma)&=&6\;\mathrm{mod}\;12,
\eea
where $N_{WC}^{k_z}(k_x,k_y)$ is the number of WCs on the $k_z$-slice at planar momentum $(k_x,k_y)$.

\section{New TCI protected by inversion and time-reversal}
\label{appn:4}

In the presence of both inversion and time-reversal, it has been shown that a double-copy of a 3D TI is still topologically nontrivial\cite{Turner2010,Hughes2011,Po2017}. Here we demonstrate that the boundary manifestation is generically a 1D helical edge mode localized on the surface of the three-disk with open boundary condition in all three directions. To see this, we first consider a 3D TI in a uniform magnetic field. Suppose that the field is along the $z$-direction, then the upper hemisphere under the field has Chern number $+1/2$ and the lower hemisphere $-1/2$, leaving a gapless equator along which a chiral edge mode runs. Now we superimpose the time-reversal copy of this resulted state, and obtain a helical edge mode along the equator.

The helical edge mode is certainly stable against any perturbation preserving time-reversal symmetry, but without additional symmetries, this circle may shrink into a point and gap out. But as long as inversion symmetry is present, this is impossible: for any gapless point, its antipodal point must also be gapless. Upon arbitrary perturbations that do not close the bulk gap, the gapless circle can be distorted to arbitrary shape as in Fig.3(d).

\section{Dimensional reduction and extension to strongly interacting SPT}
\label{appn:5}

\begin{figure}
\begin{centering}
\includegraphics[width=1\linewidth]{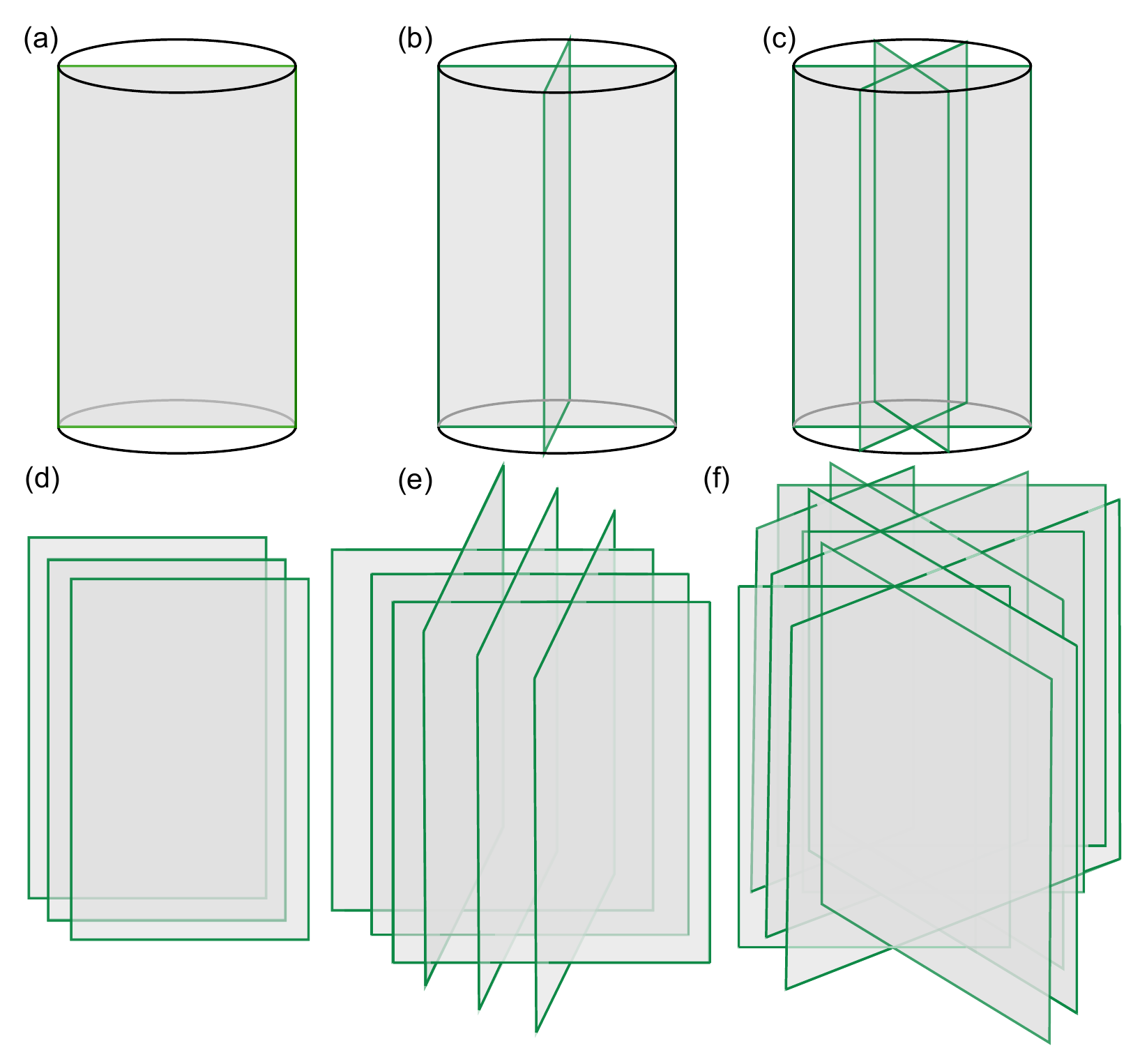}
\par\end{centering}
\protect\caption{\label{fig:4}(a-c) show in the absence of translation symmetries, the minimal constructions of twofold, fourfold and sixfold rotation symmetry protected topological states in three dimensions, where each plane is a two-dimensional internal symmetry protected topological state of either bosons or fermions. If it is a two-dimensional topological insulator, the resultant states are the topological crystalline insulators studied in the text. (d-f) are simple extensions of the minimal constructions, building three-dimensional models with translation symmetries.}
\end{figure}

In the study of mirror symmetry protected topological crystalline insulators, or more general mirror symmetry protected topological states, the perspective of dimension reduction greatly simplifies the physical picture\cite{Isobe2015,Huang2017}.
Here we show that the rotation symmetry protected topological states can be understood in this perspective.
Starting from the case of twofold rotation, we notice, from previous discussion, that the top (bottom) surface can be symmetrically gapped into two domains, leaving gapless domain walls.
And the two helical edge modes at the domain walls on the top and the bottom surface, together with the one-dimensional helical edge modes on the side surface make a closed loop, i. e., the edge of a two-dimensional topological insulator.
Therefore, the $C_2$-protected new topological crystalline insulator, when translation symmetry is ignored, is equivalent to a vertical two-dimensional topological insulator that contains the $C_2$-axis and invariant under $C_2$-rotation, schematically shown in Fig.\ref{fig:4}(a).
Similarly, one can show that the fourfold and sixfold rotation protected topological crystalline insulators are equivalent to, respectively, a set two and three vertical two-dimensional topological insulator, symmetrically arranged about the rotation axis., shown in Fig.\ref{fig:4}(b,c).
These two-dimensional topological insulators are hence the `skeletons' of the topological crystalline insulators, or their minimal constructions.
It is easy to extend these constructions to full three-dimensional models with translation symmetry: one simply replaces each two-dimensional topological insulator with a parallel array of two-dimensional topological insulators. [see Fig.\ref{fig:4}(d-f)].
For the case of fourfold rotation, this construction is essentially the same as the coupled wire construction in Ref.[\onlinecite{Song2017}], because each two-dimensional topological insulator can be constructed from helical wires.
These simple constructions can be readily generalized to strongly interacting systems of bosons and fermions.
We notice that a two-dimensional topological insulator is a fermion SPT protected by some internal symmetry that is time-reversal.
Therefore, we can replace the two-dimensional topological insulator with any two-dimensional symmetry protected topological state protected by internal symmetry group $G$\cite{Gu2009,Chen2012}, so that the resulted construction is a three-dimensional $G\times{C}_n$ symmetry protected topological state, having surface states on the top and the bottom surfaces, as well as $n$ one-dimensional edge modes on the side surface, in a cylinder geometry.
\end{appendix}
\twocolumngrid


\begin{thebibliography}{38}
\expandafter\ifx\csname natexlab\endcsname\relax\def\natexlab#1{#1}\fi
\expandafter\ifx\csname bibnamefont\endcsname\relax
  \def\bibnamefont#1{#1}\fi
\expandafter\ifx\csname bibfnamefont\endcsname\relax
  \def\bibfnamefont#1{#1}\fi
\expandafter\ifx\csname citenamefont\endcsname\relax
  \def\citenamefont#1{#1}\fi
\expandafter\ifx\csname url\endcsname\relax
  \def\url#1{\texttt{#1}}\fi
\expandafter\ifx\csname urlprefix\endcsname\relax\def\urlprefix{URL }\fi
\providecommand{\bibinfo}[2]{#2}
\providecommand{\eprint}[2][]{\url{#2}}

\bibitem[{\citenamefont{Bardeen}(1969)}]{Bardeen1969}
\bibinfo{author}{\bibfnamefont{W.~A.} \bibnamefont{Bardeen}},
  \bibinfo{journal}{Phys. Rev.} \textbf{\bibinfo{volume}{184}},
  \bibinfo{pages}{1848} (\bibinfo{year}{1969}),
  \urlprefix\url{https://link.aps.org/doi/10.1103/PhysRev.184.1848}.

\bibitem[{\citenamefont{Adler}(1969)}]{Adler1969}
\bibinfo{author}{\bibfnamefont{S.~L.} \bibnamefont{Adler}},
  \bibinfo{journal}{Phys. Rev.} \textbf{\bibinfo{volume}{177}},
  \bibinfo{pages}{2426} (\bibinfo{year}{1969}),
  \urlprefix\url{https://link.aps.org/doi/10.1103/PhysRev.177.2426}.

\bibitem[{\citenamefont{Bell and Jackiw}(1969)}]{Bell1969}
\bibinfo{author}{\bibfnamefont{J.~S.} \bibnamefont{Bell}} \bibnamefont{and}
  \bibinfo{author}{\bibfnamefont{R.}~\bibnamefont{Jackiw}},
  \bibinfo{journal}{Il Nuovo Cimento A (1965-1970)}
  \textbf{\bibinfo{volume}{60}}, \bibinfo{pages}{47} (\bibinfo{year}{1969}),
  ISSN \bibinfo{issn}{1826-9869},
  \urlprefix\url{https://doi.org/10.1007/BF02823296}.

\bibitem[{\citenamefont{Nielsen and Ninomiya}(1981)}]{Nielson1981}
\bibinfo{author}{\bibfnamefont{H.}~\bibnamefont{Nielsen}} \bibnamefont{and}
  \bibinfo{author}{\bibfnamefont{M.}~\bibnamefont{Ninomiya}},
  \bibinfo{journal}{Physics Letters B} \textbf{\bibinfo{volume}{105}},
  \bibinfo{pages}{219 } (\bibinfo{year}{1981}), ISSN \bibinfo{issn}{0370-2693},
  \urlprefix\url{http://www.sciencedirect.com/science/article/pii/0370269381910261}.

\bibitem[{\citenamefont{Niemi and Semenoff}(1983)}]{Niemi1983}
\bibinfo{author}{\bibfnamefont{A.~J.} \bibnamefont{Niemi}} \bibnamefont{and}
  \bibinfo{author}{\bibfnamefont{G.~W.} \bibnamefont{Semenoff}},
  \bibinfo{journal}{Phys. Rev. Lett.} \textbf{\bibinfo{volume}{51}},
  \bibinfo{pages}{2077} (\bibinfo{year}{1983}),
  \urlprefix\url{https://link.aps.org/doi/10.1103/PhysRevLett.51.2077}.

\bibitem[{\citenamefont{Haldane}(1988)}]{Haldane1988}
\bibinfo{author}{\bibfnamefont{F.~D.~M.} \bibnamefont{Haldane}},
  \bibinfo{journal}{Phys. Rev. Lett.} \textbf{\bibinfo{volume}{61}},
  \bibinfo{pages}{2015} (\bibinfo{year}{1988}),
  \urlprefix\url{https://link.aps.org/doi/10.1103/PhysRevLett.61.2015}.

\bibitem[{\citenamefont{Fang and Fu}(2015)}]{Fang2015}
\bibinfo{author}{\bibfnamefont{C.}~\bibnamefont{Fang}} \bibnamefont{and}
  \bibinfo{author}{\bibfnamefont{L.}~\bibnamefont{Fu}}, \bibinfo{journal}{Phys.
  Rev. B} \textbf{\bibinfo{volume}{91}}, \bibinfo{pages}{161105}
  (\bibinfo{year}{2015}),
  \urlprefix\url{https://link.aps.org/doi/10.1103/PhysRevB.91.161105}.

\bibitem[{\citenamefont{Shiozaki et~al.}(2015)\citenamefont{Shiozaki, Sato, and
  Gomi}}]{Shiozaki2015}
\bibinfo{author}{\bibfnamefont{K.}~\bibnamefont{Shiozaki}},
  \bibinfo{author}{\bibfnamefont{M.}~\bibnamefont{Sato}}, \bibnamefont{and}
  \bibinfo{author}{\bibfnamefont{K.}~\bibnamefont{Gomi}},
  \bibinfo{journal}{Phys. Rev. B} \textbf{\bibinfo{volume}{91}},
  \bibinfo{pages}{155120} (\bibinfo{year}{2015}),
  \urlprefix\url{https://link.aps.org/doi/10.1103/PhysRevB.91.155120}.

\bibitem[{\citenamefont{Zhang and Hu}(2001)}]{Zhang2001}
\bibinfo{author}{\bibfnamefont{S.-C.} \bibnamefont{Zhang}} \bibnamefont{and}
  \bibinfo{author}{\bibfnamefont{J.}~\bibnamefont{Hu}},
  \bibinfo{journal}{Science} \textbf{\bibinfo{volume}{294}},
  \bibinfo{pages}{823} (\bibinfo{year}{2001}), ISSN \bibinfo{issn}{0036-8075},
  \eprint{http://science.sciencemag.org/content/294/5543/823.full.pdf},
  \urlprefix\url{http://science.sciencemag.org/content/294/5543/823}.

\bibitem[{\citenamefont{Fu et~al.}(2007)\citenamefont{Fu, Kane, and
  Mele}}]{Fu2007}
\bibinfo{author}{\bibfnamefont{L.}~\bibnamefont{Fu}},
  \bibinfo{author}{\bibfnamefont{C.~L.} \bibnamefont{Kane}}, \bibnamefont{and}
  \bibinfo{author}{\bibfnamefont{E.~J.} \bibnamefont{Mele}},
  \bibinfo{journal}{Phys. Rev. Lett.} \textbf{\bibinfo{volume}{98}},
  \bibinfo{pages}{106803} (\bibinfo{year}{2007}),
  \urlprefix\url{https://link.aps.org/doi/10.1103/PhysRevLett.98.106803}.

\bibitem[{\citenamefont{Ando and Fu}(2015)}]{Ando2015}
\bibinfo{author}{\bibfnamefont{Y.}~\bibnamefont{Ando}} \bibnamefont{and}
  \bibinfo{author}{\bibfnamefont{L.}~\bibnamefont{Fu}},
  \bibinfo{journal}{Annual Review of Condensed Matter Physics}
  \textbf{\bibinfo{volume}{6}}, \bibinfo{pages}{361} (\bibinfo{year}{2015}),
  \eprint{https://doi.org/10.1146/annurev-conmatphys-031214-014501},
  \urlprefix\url{https://doi.org/10.1146/annurev-conmatphys-031214-014501}.

\bibitem[{\citenamefont{Chiu et~al.}(2016)\citenamefont{Chiu, Teo, Schnyder,
  and Ryu}}]{Chiu2016}
\bibinfo{author}{\bibfnamefont{C.-K.} \bibnamefont{Chiu}},
  \bibinfo{author}{\bibfnamefont{J.~C.} \bibnamefont{Teo}},
  \bibinfo{author}{\bibfnamefont{A.~P.} \bibnamefont{Schnyder}},
  \bibnamefont{and} \bibinfo{author}{\bibfnamefont{S.}~\bibnamefont{Ryu}},
  \bibinfo{journal}{Rev. Mod. Phys.} \textbf{\bibinfo{volume}{88}},
  \bibinfo{pages}{035005} (\bibinfo{year}{2016}).

\bibitem[{\citenamefont{Shiozaki et~al.}(2017)\citenamefont{Shiozaki, Sato, and
  Gomi}}]{Shiozaki2017}
\bibinfo{author}{\bibfnamefont{K.}~\bibnamefont{Shiozaki}},
  \bibinfo{author}{\bibfnamefont{M.}~\bibnamefont{Sato}}, \bibnamefont{and}
  \bibinfo{author}{\bibfnamefont{K.}~\bibnamefont{Gomi}},
  \bibinfo{journal}{Phys. Rev. B} \textbf{\bibinfo{volume}{95}},
  \bibinfo{pages}{235425} (\bibinfo{year}{2017}),
  \urlprefix\url{https://link.aps.org/doi/10.1103/PhysRevB.95.235425}.

\bibitem[{\citenamefont{Song et~al.}(2017)\citenamefont{Song, Fang, and
  Fang}}]{Song2017}
\bibinfo{author}{\bibfnamefont{Z.}~\bibnamefont{Song}},
  \bibinfo{author}{\bibfnamefont{Z.}~\bibnamefont{Fang}}, \bibnamefont{and}
  \bibinfo{author}{\bibfnamefont{C.}~\bibnamefont{Fang}},
  \bibinfo{journal}{arXiv:1708.02952}  (\bibinfo{year}{2017}).

\bibitem[{\citenamefont{Fang et~al.}(2012)\citenamefont{Fang, Gilbert, and
  Bernevig}}]{Fang2012}
\bibinfo{author}{\bibfnamefont{C.}~\bibnamefont{Fang}},
  \bibinfo{author}{\bibfnamefont{M.~J.} \bibnamefont{Gilbert}},
  \bibnamefont{and} \bibinfo{author}{\bibfnamefont{B.~A.}
  \bibnamefont{Bernevig}}, \bibinfo{journal}{Phys. Rev. B}
  \textbf{\bibinfo{volume}{86}}, \bibinfo{pages}{115112}
  (\bibinfo{year}{2012}),
  \urlprefix\url{https://link.aps.org/doi/10.1103/PhysRevB.86.115112}.

\bibitem[{\citenamefont{Fu}(2011)}]{Fu2011}
\bibinfo{author}{\bibfnamefont{L.}~\bibnamefont{Fu}}, \bibinfo{journal}{Phys.
  Rev. Lett.} \textbf{\bibinfo{volume}{106}}, \bibinfo{pages}{106802}
  (\bibinfo{year}{2011}),
  \urlprefix\url{https://link.aps.org/doi/10.1103/PhysRevLett.106.106802}.

\bibitem[{\citenamefont{Liu et~al.}(2014)\citenamefont{Liu, Zhang, and
  VanLeeuwen}}]{Liu2014}
\bibinfo{author}{\bibfnamefont{C.-X.} \bibnamefont{Liu}},
  \bibinfo{author}{\bibfnamefont{R.-X.} \bibnamefont{Zhang}}, \bibnamefont{and}
  \bibinfo{author}{\bibfnamefont{B.~K.} \bibnamefont{VanLeeuwen}},
  \bibinfo{journal}{Phys. Rev. B} \textbf{\bibinfo{volume}{90}},
  \bibinfo{pages}{085304} (\bibinfo{year}{2014}).

\bibitem[{\citenamefont{Alexandradinata
  et~al.}(2014)\citenamefont{Alexandradinata, Fang, Gilbert, and
  Bernevig}}]{Alexandradinata2014}
\bibinfo{author}{\bibfnamefont{A.}~\bibnamefont{Alexandradinata}},
  \bibinfo{author}{\bibfnamefont{C.}~\bibnamefont{Fang}},
  \bibinfo{author}{\bibfnamefont{M.~J.} \bibnamefont{Gilbert}},
  \bibnamefont{and} \bibinfo{author}{\bibfnamefont{B.~A.}
  \bibnamefont{Bernevig}}, \bibinfo{journal}{Phys. Rev. Lett.}
  \textbf{\bibinfo{volume}{113}}, \bibinfo{pages}{116403}
  (\bibinfo{year}{2014}).

\bibitem[{\citenamefont{Wieder et~al.}(2017)\citenamefont{Wieder, Bradlyn,
  Wang, Cano, Kim, Kim, Rappe, Kane, and Bernevig}}]{Wieder2017}
\bibinfo{author}{\bibfnamefont{B.~J.} \bibnamefont{Wieder}},
  \bibinfo{author}{\bibfnamefont{B.}~\bibnamefont{Bradlyn}},
  \bibinfo{author}{\bibfnamefont{Z.}~\bibnamefont{Wang}},
  \bibinfo{author}{\bibfnamefont{J.}~\bibnamefont{Cano}},
  \bibinfo{author}{\bibfnamefont{Y.}~\bibnamefont{Kim}},
  \bibinfo{author}{\bibfnamefont{H.-S.~D.} \bibnamefont{Kim}},
  \bibinfo{author}{\bibfnamefont{A.~M.} \bibnamefont{Rappe}},
  \bibinfo{author}{\bibfnamefont{C.~L.} \bibnamefont{Kane}}, \bibnamefont{and}
  \bibinfo{author}{\bibfnamefont{B.~A.} \bibnamefont{Bernevig}},
  \bibinfo{journal}{arXiv:1705.01617}  (\bibinfo{year}{2017}).

\bibitem[{\citenamefont{Wang et~al.}(2016)\citenamefont{Wang, Alexandradinata,
  Cava, and Bernevig}}]{Wang2016}
\bibinfo{author}{\bibfnamefont{Z.}~\bibnamefont{Wang}},
  \bibinfo{author}{\bibfnamefont{A.}~\bibnamefont{Alexandradinata}},
  \bibinfo{author}{\bibfnamefont{R.~J.} \bibnamefont{Cava}}, \bibnamefont{and}
  \bibinfo{author}{\bibfnamefont{B.~A.} \bibnamefont{Bernevig}},
  \bibinfo{journal}{Nature} \textbf{\bibinfo{volume}{532}},
  \bibinfo{pages}{189} (\bibinfo{year}{2016}),
  \urlprefix\url{http://dx.doi.org/10.1038/nature17410}.

\bibitem[{\citenamefont{Chang et~al.}(2017)\citenamefont{Chang, Erten, and
  Coleman}}]{Chang2017}
\bibinfo{author}{\bibfnamefont{P.-Y.} \bibnamefont{Chang}},
  \bibinfo{author}{\bibfnamefont{O.}~\bibnamefont{Erten}}, \bibnamefont{and}
  \bibinfo{author}{\bibfnamefont{P.}~\bibnamefont{Coleman}},
  \bibinfo{journal}{Nat Phys} \textbf{\bibinfo{volume}{13}},
  \bibinfo{pages}{794} (\bibinfo{year}{2017}),
  \urlprefix\url{http://dx.doi.org/10.1038/nphys4092}.

\bibitem[{\citenamefont{Shiozaki and Sato}(2014)}]{Shiozaki2014}
\bibinfo{author}{\bibfnamefont{K.}~\bibnamefont{Shiozaki}} \bibnamefont{and}
  \bibinfo{author}{\bibfnamefont{M.}~\bibnamefont{Sato}},
  \bibinfo{journal}{Phys. Rev. B} \textbf{\bibinfo{volume}{90}},
  \bibinfo{pages}{165114} (\bibinfo{year}{2014}),
  \urlprefix\url{https://link.aps.org/doi/10.1103/PhysRevB.90.165114}.

\bibitem[{\citenamefont{Benalcazar
  et~al.}(2017{\natexlab{a}})\citenamefont{Benalcazar, Bernevig, and
  Hughes}}]{Benalcazar2017}
\bibinfo{author}{\bibfnamefont{W.~A.} \bibnamefont{Benalcazar}},
  \bibinfo{author}{\bibfnamefont{B.~A.} \bibnamefont{Bernevig}},
  \bibnamefont{and} \bibinfo{author}{\bibfnamefont{T.~L.}
  \bibnamefont{Hughes}}, \bibinfo{journal}{Science}
  \textbf{\bibinfo{volume}{357}}, \bibinfo{pages}{61}
  (\bibinfo{year}{2017}{\natexlab{a}}), ISSN \bibinfo{issn}{0036-8075},
  \eprint{http://science.sciencemag.org/content/357/6346/61.full.pdf},
  \urlprefix\url{http://science.sciencemag.org/content/357/6346/61}.

\bibitem[{\citenamefont{Schindler et~al.}(2017)\citenamefont{Schindler, Cook,
  Vergniory, Wang, Parkin, Bernevig, and Neupert}}]{Schindler2017}
\bibinfo{author}{\bibfnamefont{F.}~\bibnamefont{Schindler}},
  \bibinfo{author}{\bibfnamefont{A.~M.} \bibnamefont{Cook}},
  \bibinfo{author}{\bibfnamefont{M.~G.} \bibnamefont{Vergniory}},
  \bibinfo{author}{\bibfnamefont{Z.}~\bibnamefont{Wang}},
  \bibinfo{author}{\bibfnamefont{S.~S.~P.} \bibnamefont{Parkin}},
  \bibinfo{author}{\bibfnamefont{B.~A.} \bibnamefont{Bernevig}},
  \bibnamefont{and} \bibinfo{author}{\bibfnamefont{T.}~\bibnamefont{Neupert}},
  \bibinfo{journal}{arXiv:1708.03636}  (\bibinfo{year}{2017}).

\bibitem[{\citenamefont{Langbehn et~al.}(2017)\citenamefont{Langbehn, Peng,
  Trifunovic, Oppen, and Brouwer}}]{Langbehn2017}
\bibinfo{author}{\bibfnamefont{J.}~\bibnamefont{Langbehn}},
  \bibinfo{author}{\bibfnamefont{Y.}~\bibnamefont{Peng}},
  \bibinfo{author}{\bibfnamefont{L.}~\bibnamefont{Trifunovic}},
  \bibinfo{author}{\bibfnamefont{F.~v.} \bibnamefont{Oppen}}, \bibnamefont{and}
  \bibinfo{author}{\bibfnamefont{P.~W.} \bibnamefont{Brouwer}},
  \bibinfo{journal}{arXiv:1708.03640}  (\bibinfo{year}{2017}).

\bibitem[{\citenamefont{Benalcazar
  et~al.}(2017{\natexlab{b}})\citenamefont{Benalcazar, Bernevig, and
  Hughes}}]{Benalcazar2017a}
\bibinfo{author}{\bibfnamefont{W.~A.} \bibnamefont{Benalcazar}},
  \bibinfo{author}{\bibfnamefont{B.~A.} \bibnamefont{Bernevig}},
  \bibnamefont{and} \bibinfo{author}{\bibfnamefont{T.~L.}
  \bibnamefont{Hughes}}, \bibinfo{journal}{arXiv:1708.04230}
  (\bibinfo{year}{2017}{\natexlab{b}}).

\bibitem[{\citenamefont{Turner et~al.}(2010)\citenamefont{Turner, Zhang, and
  Vishwanath}}]{Turner2010}
\bibinfo{author}{\bibfnamefont{A.~M.} \bibnamefont{Turner}},
  \bibinfo{author}{\bibfnamefont{Y.}~\bibnamefont{Zhang}}, \bibnamefont{and}
  \bibinfo{author}{\bibfnamefont{A.}~\bibnamefont{Vishwanath}},
  \bibinfo{journal}{Phys. Rev. B} \textbf{\bibinfo{volume}{82}},
  \bibinfo{pages}{241102} (\bibinfo{year}{2010}),
  \urlprefix\url{https://link.aps.org/doi/10.1103/PhysRevB.82.241102}.

\bibitem[{\citenamefont{Hughes et~al.}(2011)\citenamefont{Hughes, Prodan, and
  Bernevig}}]{Hughes2011}
\bibinfo{author}{\bibfnamefont{T.~L.} \bibnamefont{Hughes}},
  \bibinfo{author}{\bibfnamefont{E.}~\bibnamefont{Prodan}}, \bibnamefont{and}
  \bibinfo{author}{\bibfnamefont{B.~A.} \bibnamefont{Bernevig}},
  \bibinfo{journal}{Phys. Rev. B} \textbf{\bibinfo{volume}{83}},
  \bibinfo{pages}{245132} (\bibinfo{year}{2011}),
  \urlprefix\url{https://link.aps.org/doi/10.1103/PhysRevB.83.245132}.

\bibitem[{\citenamefont{Isobe and Fu}(2015)}]{Isobe2015}
\bibinfo{author}{\bibfnamefont{H.}~\bibnamefont{Isobe}} \bibnamefont{and}
  \bibinfo{author}{\bibfnamefont{L.}~\bibnamefont{Fu}}, \bibinfo{journal}{Phys.
  Rev. B} \textbf{\bibinfo{volume}{92}}, \bibinfo{pages}{081304}
  (\bibinfo{year}{2015}).

\bibitem[{\citenamefont{Ezawa}(2016)}]{Ezawa2016}
\bibinfo{author}{\bibfnamefont{M.}~\bibnamefont{Ezawa}},
  \bibinfo{journal}{Phys. Rev. B} \textbf{\bibinfo{volume}{94}},
  \bibinfo{pages}{155148} (\bibinfo{year}{2016}),
  \urlprefix\url{https://link.aps.org/doi/10.1103/PhysRevB.94.155148}.

\bibitem[{\citenamefont{{Huang} et~al.}(2017)\citenamefont{{Huang}, {Song},
  {Huang}, and {Hermele}}}]{Huang2017}
\bibinfo{author}{\bibfnamefont{S.-J.} \bibnamefont{{Huang}}},
  \bibinfo{author}{\bibfnamefont{H.}~\bibnamefont{{Song}}},
  \bibinfo{author}{\bibfnamefont{Y.-P.} \bibnamefont{{Huang}}},
  \bibnamefont{and}
  \bibinfo{author}{\bibfnamefont{M.}~\bibnamefont{{Hermele}}},
  \bibinfo{journal}{ArXiv e-prints}  (\bibinfo{year}{2017}),
  \eprint{1705.09243}.

\bibitem[{\citenamefont{Liu et~al.}(2013)\citenamefont{Liu, Duan, and
  Fu}}]{Liu2013}
\bibinfo{author}{\bibfnamefont{J.}~\bibnamefont{Liu}},
  \bibinfo{author}{\bibfnamefont{W.}~\bibnamefont{Duan}}, \bibnamefont{and}
  \bibinfo{author}{\bibfnamefont{L.}~\bibnamefont{Fu}}, \bibinfo{journal}{Phys.
  Rev. B} \textbf{\bibinfo{volume}{88}}, \bibinfo{pages}{241303}
  (\bibinfo{year}{2013}),
  \urlprefix\url{https://link.aps.org/doi/10.1103/PhysRevB.88.241303}.

\bibitem[{\citenamefont{Hsieh et~al.}(2012)\citenamefont{Hsieh, Lin, Liu, Duan,
  Bansil, and Fu}}]{Hsieh2012}
\bibinfo{author}{\bibfnamefont{T.~H.} \bibnamefont{Hsieh}},
  \bibinfo{author}{\bibfnamefont{H.}~\bibnamefont{Lin}},
  \bibinfo{author}{\bibfnamefont{J.}~\bibnamefont{Liu}},
  \bibinfo{author}{\bibfnamefont{W.}~\bibnamefont{Duan}},
  \bibinfo{author}{\bibfnamefont{A.}~\bibnamefont{Bansil}}, \bibnamefont{and}
  \bibinfo{author}{\bibfnamefont{L.}~\bibnamefont{Fu}},
  \textbf{\bibinfo{volume}{3}}, \bibinfo{pages}{982 EP }
  (\bibinfo{year}{2012}), \urlprefix\url{http://dx.doi.org/10.1038/ncomms1969}.

\bibitem[{\citenamefont{Hsieh et~al.}(2014)\citenamefont{Hsieh, Liu, and
  Fu}}]{Hsieh2014}
\bibinfo{author}{\bibfnamefont{T.~H.} \bibnamefont{Hsieh}},
  \bibinfo{author}{\bibfnamefont{J.}~\bibnamefont{Liu}}, \bibnamefont{and}
  \bibinfo{author}{\bibfnamefont{L.}~\bibnamefont{Fu}}, \bibinfo{journal}{Phys.
  Rev. B} \textbf{\bibinfo{volume}{92}}, \bibinfo{pages}{081112}
  (\bibinfo{year}{2014}).

\bibitem[{\citenamefont{Marzari and Vanderbilt}(1997)}]{Marzari1997}
\bibinfo{author}{\bibfnamefont{N.}~\bibnamefont{Marzari}} \bibnamefont{and}
  \bibinfo{author}{\bibfnamefont{D.}~\bibnamefont{Vanderbilt}},
  \bibinfo{journal}{Phys. Rev. B} \textbf{\bibinfo{volume}{56}},
  \bibinfo{pages}{12847} (\bibinfo{year}{1997}),
  \urlprefix\url{https://link.aps.org/doi/10.1103/PhysRevB.56.12847}.

\bibitem[{\citenamefont{Po et~al.}(2017)\citenamefont{Po, Vishwanath, and
  Watanabe}}]{Po2017}
\bibinfo{author}{\bibfnamefont{H.~C.} \bibnamefont{Po}},
  \bibinfo{author}{\bibfnamefont{A.}~\bibnamefont{Vishwanath}},
  \bibnamefont{and} \bibinfo{author}{\bibfnamefont{H.}~\bibnamefont{Watanabe}},
  \bibinfo{journal}{Nature Communications} \textbf{\bibinfo{volume}{8}},
  \bibinfo{pages}{50} (\bibinfo{year}{2017}),
  \urlprefix\url{https://doi.org/10.1038/s41467-017-00133-2}.

\bibitem[{\citenamefont{Gu and Wen}(2009)}]{Gu2009}
\bibinfo{author}{\bibfnamefont{Z.-C.} \bibnamefont{Gu}} \bibnamefont{and}
  \bibinfo{author}{\bibfnamefont{X.-G.} \bibnamefont{Wen}},
  \bibinfo{journal}{Phys. Rev. B} \textbf{\bibinfo{volume}{80}},
  \bibinfo{pages}{155131} (\bibinfo{year}{2009}),
  \urlprefix\url{https://link.aps.org/doi/10.1103/PhysRevB.80.155131}.

\bibitem[{\citenamefont{Chen et~al.}(2012)\citenamefont{Chen, Gu, Liu, and
  Wen}}]{Chen2012}
\bibinfo{author}{\bibfnamefont{X.}~\bibnamefont{Chen}},
  \bibinfo{author}{\bibfnamefont{Z.-C.} \bibnamefont{Gu}},
  \bibinfo{author}{\bibfnamefont{Z.-X.} \bibnamefont{Liu}}, \bibnamefont{and}
  \bibinfo{author}{\bibfnamefont{X.-G.} \bibnamefont{Wen}},
  \bibinfo{journal}{Science} \textbf{\bibinfo{volume}{338}},
  \bibinfo{pages}{1604} (\bibinfo{year}{2012}), ISSN \bibinfo{issn}{0036-8075},
  \eprint{http://science.sciencemag.org/content/338/6114/1604.full.pdf},
  \urlprefix\url{http://science.sciencemag.org/content/338/6114/1604}.

\end{thebibliography}
\end{document}